# Mechanisms for directed self-assembly of heteroepitaxial Ge/Si quantum dots with deterministic placement and sub-23nm spacing on SiC nanotemplates

*Christopher W. Petz[1], Dongyue Yang[2], Jeremy Levy[2], Jerrold A. Floro[1]*

[1]Department of Materials Science and Engineering, University of Virginia, Charlottesville, VA 22904

[2]Department of Physics and Astronomy, University of Pittsburgh, Pittsburgh, PA 15260

## Abstract:

Artificially ordered Ge quantum dot (QD) arrays, where confined carriers can interact via exchange coupling, may create unique functionalities such as cluster qubits and spintronic bandgap systems. Development of such arrays for quantum computing requires fine control over QD size and spatial arrangement on the sub-35 nm length scale. We employ fine-probe electron-beam irradiation to locally decompose ambient hydrocarbons onto a bare Si (001) surface. These carbonaceous patterns are annealed in UHV, forming ordered arrays of nanoscale SiC precipitates that serve as templates for subsequent Ge quantum dot self-assembly during heteroepitaxy. This templating approach has so far produced interdot spacings down to 22.5 nm, and smaller spacings should be possible. We investigate the templated feature evolution during UHV processing to identify key mechanisms that must be controlled in order to preserve pattern fidelity and reduce broadening of the quantum dot size distribution. A key finding is that the presence of a small background of excess carbon reduces Ge surface diffusion, thereby suppressing coarsening to relatively high temperatures. In fact, coarsening of the carbonaceous nanodot template prior to conversion to SiC can be a more important contributor to size dispersion, and must be avoided through control of thermal budget.

**Introduction:**

New device paradigms utilizing logic based on spin interactions in confined systems have gained considerable interest for the exploration of the basic physics of quantum structures and of new computing architectures. For example, a linear chain containing an odd number of antiferromagnetically coupled spins would act as an $S = ½$ cluster qubit, whose behavior would mirror that of a standard, single spin $s = ½$ qubit.[1,2] The advantage of a cluster qubit is that there is greatly reduced sensitivity to the detailed coupling between spins inside the cluster, and control of magnetic fields to minimize decoherence is now necessary only on the scale of the cluster length, i.e., about an order of magnitude increase in length scale for a 9 spin linear cluster. Similar considerations apply to two- or- three- dimensional spin clusters with an uncompensated spin. [1,2] When large arrays of qubits are created, artificial spintronic band gaps can be realized with electrical properties that depend purely on the geometrical arrangement[3].

While cluster qubits have the advantage of being much less sensitive to the intracluster exchange, it is still necessary to stably position spins sufficiently closely to develop exchange coupling energies comparable to $k_BT$. Recently, Pryor, et al., predicted observable exchange coupling behavior between electrons localized by adjacent heteroepitaxial Ge quantum dots embedded in Si[4]. It was shown that strain-induced changes in the band edges of Ge and Si led to formation of shallow minima in the Si conduction band above and below the Ge QDs that would confine electrons. Confinement was sufficiently weak that wavefunction decay lengths were of order 10 nm. It was shown that for adjacent

Ge QDs separated by 24 nm, the exchange interaction energy could be as large as 1.3 meV. In this paper we report our progress and understanding of a simple methodology to reliably direct epitaxial self-assembly of Ge QDs at near-20 nm length scales.

More generally, the ability to control the QD spatial positioning and size distribution represents a key enabler for any device exploiting interactions between proximal, periodic arrays of QDs. This can be achieved in principle via directed self-assembly methods that strongly constrain QD nucleation and growth. A method developed in earlier work by Guise, et al., showed that Ge quantum dots preferentially self-assembled on nanoscale silicon carbide templates, which were created by electron beam induced deposition (EBID)[5] of carbon nanodots, followed by thermal conversion to SiC. Guise, et al., achieved resultant interdot spacings down to 35 nm.[5] Using the same approach, with additional control over the processing and growth, we have further reduced the interdot spacing down to 22.5 nm, at least in small arrays, as shown in Fig. 1, where the "6" of the "nanodice" pattern is 45 nm across its long edge. The nanodice structure was fabricated by our directed self-assembly process to enable systematic transport measurements where different numbers of quantum dots are interacting. In this paper we discuss several key process parameters and mechanisms controlling the resultant QD size distribution.

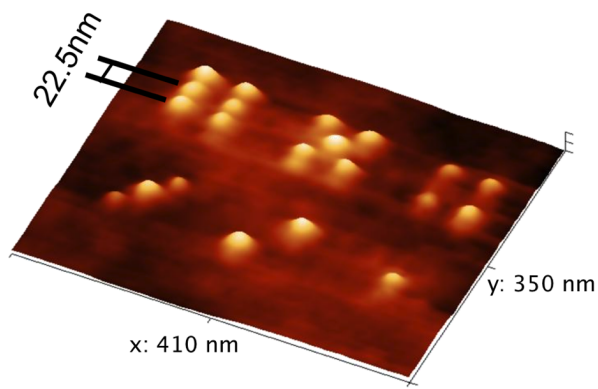

**Figure 1:** "Nano-dice" array of Ge/SiC/Si (001) quantum dots.

**Experiment:**

Si wafers with a miscut of 0.1° were cleaned via a modified IMEC/Shiraki process and passivated in the final step by oxidation in a UV-ozone environment. The $SiO_2$ layer was then stripped in a dilute HF solution immediately before loading into a Raith 150 EBL system ($P_{base}$~$10^{-7}$T) operating with an accelerating voltage of 20 keV and emission current of 21 pA. Focused electron beam irradiation of ambient hydrocarbon adsorbates on the Si(001) surface yielded islands of carbonaceous material, that we'll refer to as $C_xH_y$[5,6]. Square arrays of NxN nanodots were written, with interdot spacings of 100, 50 and 35 nm, and N as large as 300. Previous studies by Guise, et al. have characterized the size of these islands as a function of exposure time[7] and for this study we used an optimized exposure time of 3-6 ms. Next, the templated $C_xH_y$ nanodots were exposed to UV-ozone to eliminate excess hydrocarbon contamination between the patterned sites. Since this also simultaneously etched the patterned nanodots, UV-ozone exposure was critically controlled to prevent their complete removal. The patterned wafers were then inserted in to the ultra-high vacuum (UHV) molecular beam epitaxy (MBE) chamber ($P_{base}$=1x$10^{-10}$ Torr) and radiatively heated during an overnight temperature ramp to 500°C. All samples dwelt at 500°C for at least an hour prior to oxide desorption at 780°C. Throughout the deoxidation process, the surface structure was monitored with reflection high energy electron diffraction (RHEED) to ensure that a smooth, 2×1 reconstructed surface developed, as indicated by the presence of a Laue ring of diffraction spots. During the oxide desorption at 780°C, the $C_xH_y$ nanodots are believed to be fully converted to SiC, but this is discussed further below. Upon cooling to the growth temperature, 1.3 ML of Ge was deposited via magnetron sputtering in 3 mTorr of getter-purified Ar at a rate of 0.1 Å/s. In these experiments, we examine the effect of two differing growth strategies: direct deposition at 600°C with no further annealing, and deposition at 400°C with subsequent annealing at 700°C.

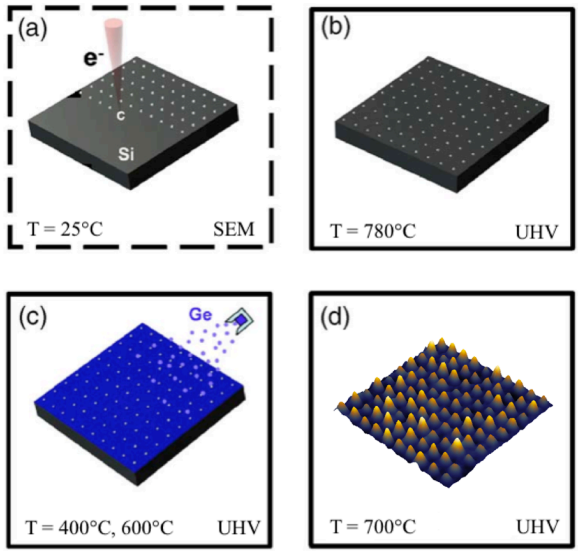

**Figure 2:** Schematic for templated Ge island growth. (a) EBID of carbonaceous template; (b) annealing in UHV for SiC transformation and desorption of $SiO_2$; (c) deposition of Ge; and (d) optional annealing of Ge/SiC islands.

Ex-situ atomic force microscopy (AFM) was performed with an NT-MDT Solver Pro-M in semi-contact mode using NSG01 tips with radius < 7 nm. Obtaining accurate topographic measurements of nanoscale islands is non-trivial when the feature radius is near that of the AFM tip. At this small size limit, the extent of tip-convolution is predominantly a function of island aspect ratio[8]. In this study, we measure the full shape of Ge QDs with diameters (Ø) between 5-25 nm and aspect ratios of 0.05-0.15. Simple geometrical calculations confirm that the *maximum* volume overestimation due to tip convolution is +6.5% for QDs with an aspect ratio of 0.15. We neglect the effects of tip convolution in all measurements reported here; however for features with an aspect ratio > 0.20, a correction factor may be necessary.

Measurements of the QD and template volumes via AFM utilizes the "flooding" procedure, where a flood plane is determined "by eye" for the flattened images, which ideally would mask any background roughness, so that only the raised volume of the templated sites is obtained. However, due to the small size of our features, even a ~5Å RMS background roughness makes it impossible to choose a single universal flood plane for an array of templated sites. We thus chose a minimum initial flood plane which served as a local peak-finder. From this point, the data masks may be enlarged pixel-by-pixel until the local flood plane reaches ~10% of the maximum height. Since the average island height is 1nm, we can select a proper flood plane within 1Å of the island base. This translates to a volume estimate error which is less than the contribution of tip convolution discussed above.

An important aspect of this work is to obtain the distribution of Ge and SiC volumes at the templated sites as a function of process conditions. We thoroughly characterized the morphology after Ge growth on the template using AFM. Next we performed chemical etching to selectively remove the Ge, using $H_2O$ and $H_2O_2$, and then measured the morphology again on the identical area of the template, giving a direct comparison of volume evolution before and after Ge growth. In some cases an HF etch was performed instead, with only minor differences in the results.

## Results and Discussion:

For device applications involving interactions between quantum dots, such as exchange coupling or tunneling, it is critically important to control the variability in size and spacing of the composite Ge/SiC/Si(001) patterned nanostructures. While control of QD *spacing* is predominantly limited by the resolution of the EBID process, the QD *size* depends on all subsequent process steps. Two aspects in this process are of key importance – the UV-ozone clean and the detailed thermal budget seen by the template. In the work reported here, we kept the UV-ozone exposure constant, but varied the thermal budget. In particular, a relatively low-temperature dwell at 500°C, which was used to drive off adsorbates from the sample prior to desorption of the passive oxide at 780°C, had a surprisingly strong effect on the final size, and the distribution of sizes, of the templated features.

Figure 3 shows AFM images of a small portion of the Ge/SiC arrays after growth at 600°C, and after growth at 400°C followed by a 700°C anneal. From such images, we obtain the distribution of feature volumes (where "feature" = SiC nanodots or Ge/SiC composite structures) through flooding analysis. Figure 4(a) shows the average volume of the SiC arrays with 35, 50, and 100nm spacings. The sample with "zero" annealing time represents the $C_xH_y$ nanodot mean volume prior to any heating in UHV. The other three samples had an identical *in vacuo* thermal budget, except for the dwell times at 500°C prior to oxide desorption. The arrays with dwell times of 2.75 and 5.0 hours were overgrown with 1.3ML of Ge at 600°C which was selectively etched off as indicated. The dwell time of 6.25 hrs had no Ge grown on the SiC template. Figure 4(a) clearly shows that the overall heating and conversion to SiC results in up to 86% reduction in feature volume from the original $C_xH_y$ nanodots. While some of this reduction is likely associated with the high temperature conversion to SiC, there is a strong dependence on the 500°C dwell time. For example, from Fig. 4(a), the mean SiC nanodot volume after 6.25 hrs dwell is reduced by 65% compared with the 2.75 hr dwell. Hence significant mass was lost during the dwell. Furthermore, as shown in Fig. 4(b), the width of the distribution of volumes increases with the dwell time at 500°C, on both an absolute- and log- scale, negatively impacting array uniformity.

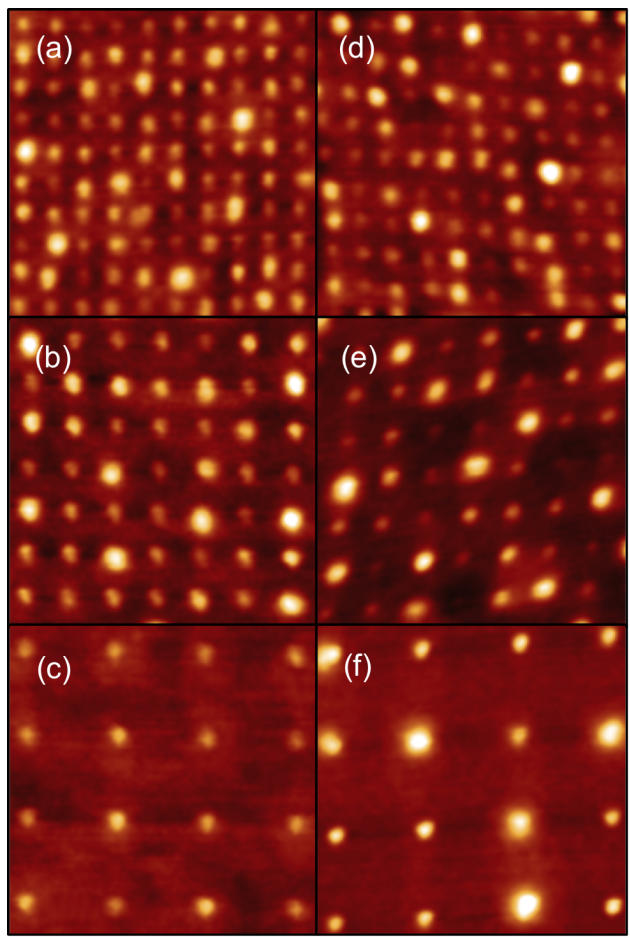

**Figure 3:** 35nm, 50nm, and 100nm arrays after (a-c) 600°C and (d-f) 400°C + 700°C depositions. Each tile is a 350nm x 350nm section from larger arrays. (Note: Small deviations in orthogonality are due to drift during AFM scanning.)

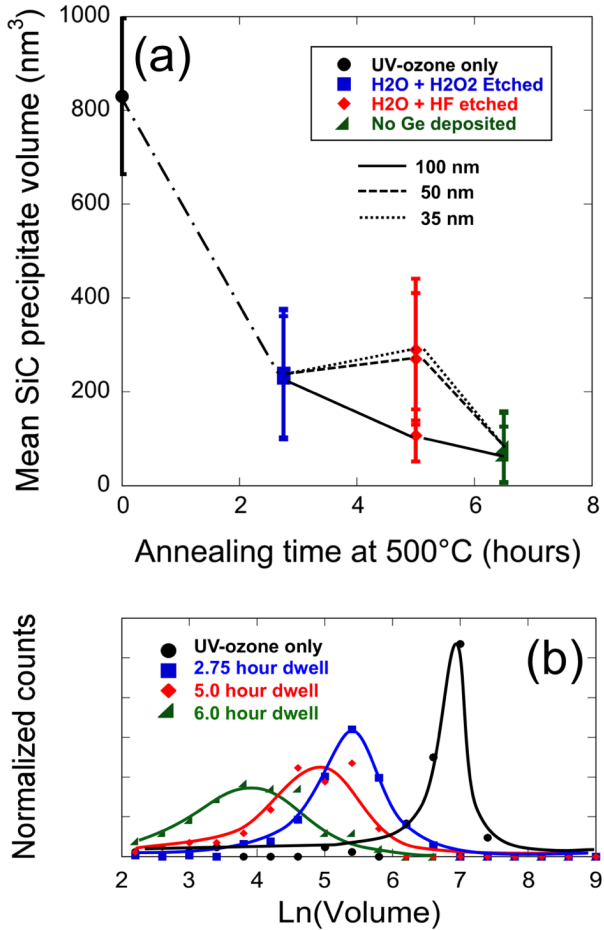

**Figure 4:** (a) Average feature volume vs. dwell times at 500°C and (b) distribution of feature volumes for the 100 nm pattern spacing for the various dwell times.

The effect of the Ge deposition parameters on the size uniformity is shown in Fig. 5. Each volume distribution is obtained from more than 200 features per array. In all cases, the red and blue curves are taken from the exact same QD arrays with Ge grown at 600°C (red) and after etching away the Ge with $H_2O$ and $H_2O_2$ (blue). Growth of 1.3 ML of Ge at 600°C results in a volume distribution that is identical to that of the underlying SiC template (obtained by selective etching). This observation holds for all three interdot spacings. On the other hand, growth of Ge at 400°C, followed by a 700°C, 30 min anneal, clearly leads to coarsening of Ge. Interestingly, the degree of coarsening *increases* with the interdot spacing. At 100 nm, a clearly bimodal distribution is produced. The bimodality is less pronounced at 50 nm spacing, and only a shoulder is observed at 35 nm.

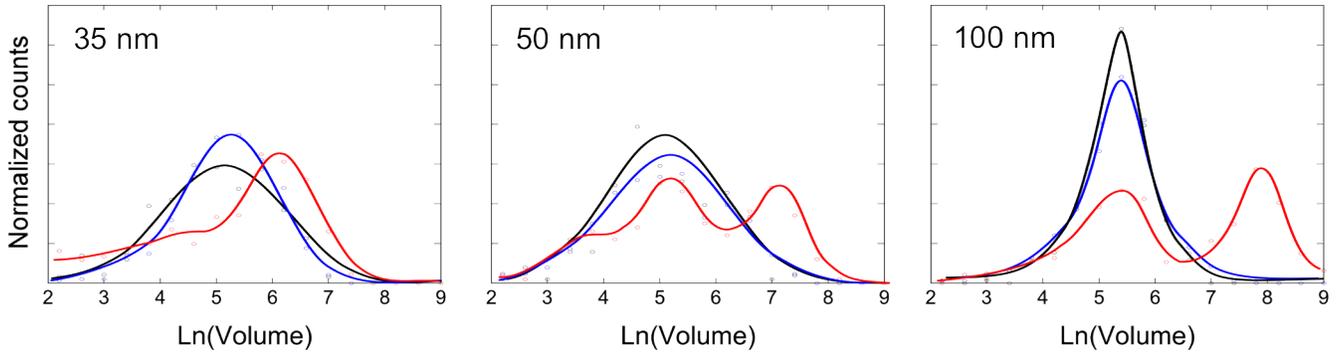

**Figure 5:** Volume distributions of Ge/SiC composite islands (2.75 hr dwell at 500°C) for two different growth strategies: (blue) Ge deposition at 600°C and (red) Ge deposition at 400°C with post annealing at 700°C. (a) 35nm spacing, (b) 50nm spacing, and (c) 100nm spacing. For comparison, SiC arrays, after Ge etching, are also shown (black).

In Fig. 6(a), we show a 35nm array after Ge deposition at 600°C along with a fast fourier transform (FFT) of the same array. By extracting the FWHM of the first order <01> peak, we determine the average spatial variation to be ±1.6nm. While positional accuracy of these QDs is adequate for device needs, it is more difficult to control the diameter of the SiC template. Figure 6(b) is a cumulative probability plot of QD diameters for the same 35nm array. In the observed distribution 51% of islands have FWHM ≤10nm and 95% are ≤15nm.

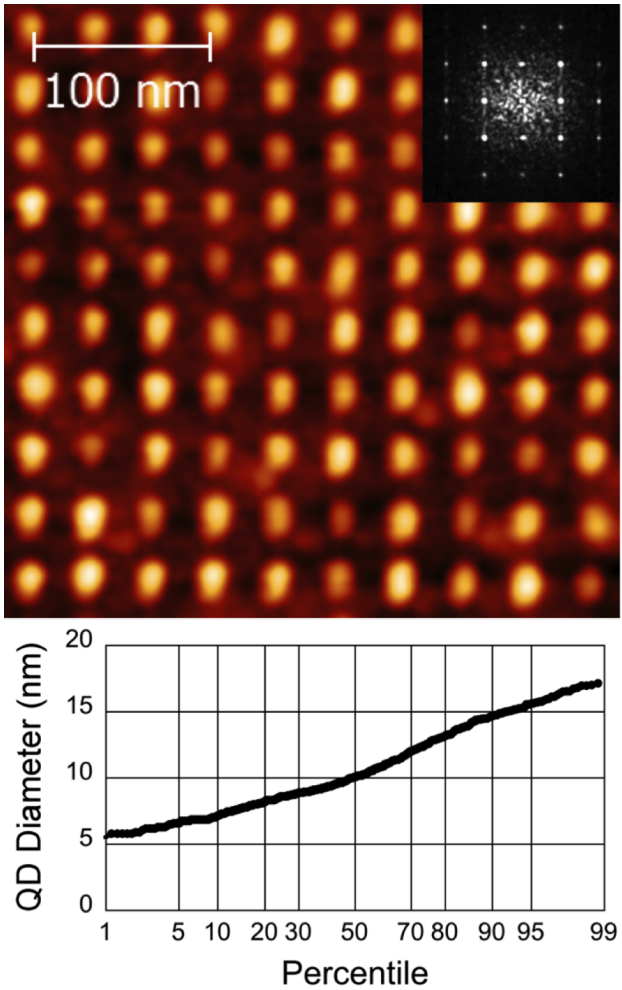

**Figure 6:** (a) AFM image of a 10x10 array section of composite Ge/SiC QDs with 35nm spacing. The inset displays a fast fourier transform of the full array; the FWHM of the <10> peak determines a positional error of ±1.6nm. (b) Cumulative probability plot of the QD diameters for the same 35nm spacing array. 51% of islands have Ø ≤10 nm, while 95% have Ø ≤15 nm.

Fig. 7 shows selected profiles of SiC nanoprecipitates (black line) and Ge/SiC composite QDs from a 100nm array after *in situ* annealing at 700°C. The typical profile shape is roughly Gaussian since nanoscale faceting at this size scale cannot be resolved with AFM. Instead we characterize the profiles by the maximum observed sidewall angle. SiC nanoprecipitates exhibit 7-10° angles while the composite Ge/SiC islands vary from 10-17°.

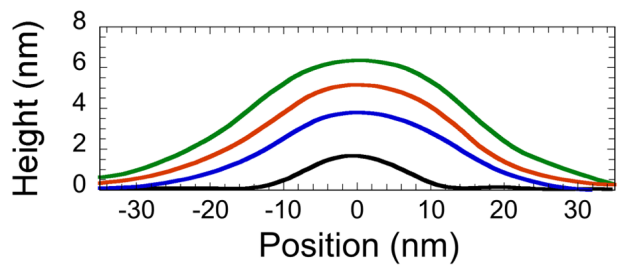

**Figure 7:** Line profiles of a typical SiC nanoprecipitate (black line) and coarsened Ge/SiC composite QDs (blue, red, green lines) from 100nm spacing arrays.

We have further examined and refined the process of EBID patterning of Ge QDs originally developed by Guise, et al., where focused electron beam-induced cracking of native adsorbed hydrocarbons on the Si surface is used to write patterns of nanoscale carbonaceous, $C_xH_y$, nanodots, that, when converted to SiC, act to template subsequent Ge growth. Our results show that: (1) the relatively narrow size distribution provided by the EBID write process can be lost during moderate temperature heating dwells; (2) 1.3 ML Ge deposited at 600°C on the pattern is conformal, coating both the Si and the SiC nanodots uniformly; (3) upon annealing to higher temperatures, Ge can undergo coarsening – while the Ge dots still reside on the templated pattern sites, their size distribution is greatly broadened. In the following, we discuss the mechanisms and implications of these results.

### *SiC Template Formation*

During heating of the EBID-written $C_xH_y$ patterned features in UHV to a maximum temperature of 780°C, a 500°C dwell is found to strongly affect the final feature size, with longer dwells leading to much smaller features, and a much broader size distribution. Possible mechanisms for these changes include evaporative loss, diffusive loss into the substrate, and interparticle coarsening. We eliminate carbon diffusion into Si as a loss mechanism, since it is not significant below 600°C.[9,10] Interdot coarsening by itself cannot lead to loss of mass. Hence we speculate that at 500°C, volatile components of the $C_xH_y$ nanodots evaporate, leaving behind a carbon-enriched feature; others have similarly reported complete dehydrogenation of $C_xH_y$ between 450-550°C[6]. The concomitant broadening of the size distribution either implies the evaporative loss process is locally non-uniform, or that coarsening is occurring simultaneously with evaporation, perhaps mediated by a non-volatile surface diffusing species, e.g., carbon atoms.

During heating to 780°C to desorb the oxide, we have invoked here that the nanodots convert to crystalline SiC. Guise, et al., found that macro-area EBID (300 um diameter region vs. our 20 nm diameter features) began converting to SiC above about 600°C, and were fully converted to SiC at about 900°C.[6] Although our maximum temperature is lower, the amount of deposited carbon per area in our experiments is also much lower than in Guise, et al., so much less interdiffusion is required to complete the carbidization. Fig. 5 also shows distribution broadening as a function of feature spacing suggesting that carbide coarsening is enhanced for the 50nm and 35nm arrays. Though $C_xH_y$ coarsening is limited at 500°C, it is likely that C and/or SiC coarsening continues throughout the entire carbidization and oxide desorption process. This further indicates that control of the entire thermal budget is crucial for retention of a narrow distribution.

We do not know yet what polymorph of SiC is formed, or whether these nanoscale precipitates are in an epitaxial relationship with the Si substrate. While there appears to be great variability of reported structures in the literature, CVD-supplied carbon on clean Si has been observed to result in formation of the cubic polymorph, 3C-SiC, in a cube-on-cube epitaxial relationship with the Si[11-13]. Although the carbides are epitaxial with the Si, their interface is incommensurate, unsurprising given the 20% lattice mismatch.

Our work suggests that minimizing the time-temperature product prior to the conversion of the $C_xH_y$ nanodots to SiC is crucial to keeping the narrowest possible distribution in the templated array. This is of key importance, since we ultimately find that the Ge size distribution can never be better that that of the template.

### *Ge directed self-assembly*

Epitaxial self-assembly of Ge QDs on unpatterned Si (001) at 600°C results in coherent islands in the so-called[14] pyramid and dome morphologies, as shown in Fig. 8. During the growth of this film, the first 3-4 ML of Ge grows as a thermodynamically stable planar wetting layer, and all Ge deposition beyond this critical thickness forms the QDs. Clearly this process results in QDs with random spatial arrangements. In order to obtain deterministic placement of dots, directed self-assembly is required.

Our goal in this work, building on previous results[5], was to refine and better understand the carbide-based methodology for directing self-assembly of 3D Ge quantum dots. Fig. 5 shows that growth of 1.3 ML at 600°C results in a volume distribution that exactly replicates the underlying SiC nanodot distribution, implying that we actually obtain conformal layer growth of Ge over both the SiC nanodots and the bare Si regions in between. Hence, *morphological* Ge QDs do not form, although we argue below that *electronic* QDs may still form. Since our Ge layer is well below the critical wetting layer thickness of 3-4 ML, it may seem unsurprising that morphological QDs are not observed. However, for the same layer thickness, we note that 3D Ge quantum dots *do* form in the case where a sufficiently high annealing temperature (here, 700°C) is used. Note that this would not be the case for unpatterned Si – the wetting layer is an equilibrium feature. Furthermore, at 700°C the morphological Ge QDs clearly localize on the SiC patterned sites, demonstrating that the carbides do preferentially attract Ge, implying that a local reduction in the critical thickness for QD formation is obtained, consistent with other results[5,15,16].

Analysis of Fig. 8 indicates that for growth of Ge QDs on unpatterned Si at 600°C, the mean interdot spacing is ≈ 120 nm. Hence adatom diffusion lengths must be about 60 nm. Since only conformal Ge growth was observed at 600°C on the pattern, even for feature spacings of only 35 nm, this suggests that surface diffusion of Ge is significantly inhibited in the pattern region. This is likely due to the retention of some spurious, dilute carbon coverage on the Si regions between the SiC sites. Carbon is known to suppress Ge surface diffusion on Si (001)[17].

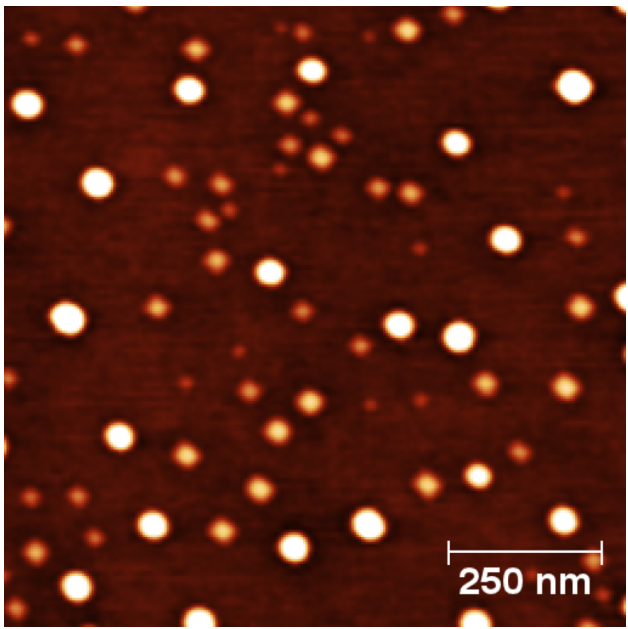

**Figure 8:** Self-assembled Ge QDs on unpatterned Si. 6ML of Ge was deposited at 600°C at a rate of 0.1Å/s.

For Ge post-annealed at 700°C, surface diffusion rates are now sufficiently large that coarsening is able to occur. We obtain an estimate of the capture zone radius of Ge for each distribution as:

$$R = \sqrt{\frac{\overline{V}_{comp} - \overline{V}_{SiC}}{\pi(h_{dep} - h_{WL})}}$$

where $\overline{V}_{comp}$ is the mean volume of the Ge/SiC composite structure, $\overline{V}_{SiC}$ is the mean volume of the carbide template, $h_{dep}$ = 1.3 ML is the total deposited Ge thickness, and $h_{WL}$ is the thickness of any retained wetting layer on the Si surface between the template sites. We have assumed a circular capture zone, and that the area of the template sites is small compared to the area in between. The mean volumes are obtained directly from the appropriate distributions in Fig. 5. We then calculate L for two different assumptions, $h_{WL}$ = 0 (i.e., Si can be completely scavenged from the adjacent bare substrate onto the SiC sites), and $h_{WL}$ = 1 (i.e., for surface energy minimization, one ML of Ge is always retained on the bare Si).

Fig. 9 shows the result of this analysis. The assumption of $h_{WL}$ = 0 implies that, for the 100 nm array, the diffusion length would be less than the nearest-neighbor spacing, i.e., each site would only scavenge Ge from the wetting layer, and not from neighboring sites. Visual inspection of the AFM images (see Fig. 3) and the strongly bimodal distribution clearly show that this is incorrect. Numerous template sites in the 100 nm array appear to be completely scavenged of their Ge by dots at least 2 sites away. Hence, the assumption of $h_{WL}$ = 1 ML, the blue line on Fig. 9, provides a more reasonable agreement with the data.

The analysis in Fig. 9 also shows that the capture zone radius decreases with decreasing array spacing, rather than remaining constant as might be expected on a monolithic Si surface. As mentioned above, an earlier STM study by Leifeld has shown that C-alloying of a Si(001) surface can modify the reconstructed surface and severely limit adatom diffusion[17]. The reduced diffusion lengths observed here with decreasing array spacing are consistent with limited diffusion associated with C surface modification. In the high-density patterned regions (50nm and 35nm), off-site C contamination is likely to be larger. That being said, C-limited kinetics ultimately minimizes coarsening processes and limits the maximum QD volume.

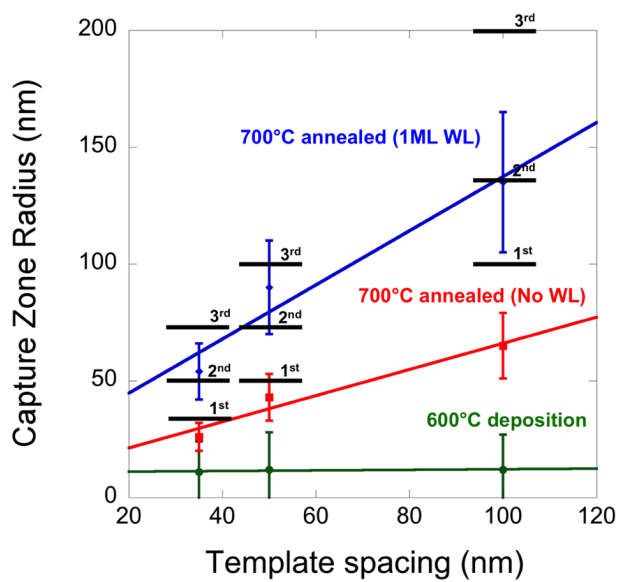

**Figure 9:** Calculated capture zone radius for templated, self-assembled Ge QDs in the two different growth schemes presented here. The red line assumes the SiC nanoprecipitates scavenge all Ge within the capture zone radius. The blue line shows diffusion radii assuming a 1ML Ge wetting layer remains between templated spots. Distances to the 1$^{st}$-3$^{rd}$ nearest neighboring islands are marked with black lines.

Selected profiles of the largest coarsened islands are shown in Fig. 8. We compare the shape of coarsened Ge islands to that of bare SiC nanoprecipitates to examine the extent of Ge self-assembly during island formation. Clearly, control over island size is lost when long range diffusion activates. Furthermore, with Ge accumulation we observe average sidewall angles ranging from 11-17°C. 3D analyses of the largest dots do not reveal the four-fold symmetric shape of compressively strained Ge {105} facets. Instead, we find a continuum of angles and Ge islands have a nearly round shape. This suggests that if the Ge is epitaxial, it may inherit more complex faceting from the underlying SiC. Determination of the exact faceting behavior of these morphological Ge QDs is thus difficult due to the small island size.

*<u>Device implications</u>*

Per the theoretical calculations of Pryor, et al.[4], tensile-strained Si above a Ge QD can trap electrons in localized energy states in the Si just above the apex of the dot, due to strain-induced reduction of the conduction band energy. Electrons inhabiting identical states in adjacent quantum dots can undergo coupling via exchange interactions, with energies of order 1 meV for interdot separations (center-to-center) of order 25 nm. Clearly this also sets strict limitations on dot size. We are achieving the requisite length scales, down to 22.5 nm interdot spacing (Fig. 1) and 10nm QD widths (Fig. 6(b)) through the use of our carbide-based directed self-assembly. Since the Ge is conformal when coarsening is avoided, it is certainly possible to write even finer spacings using EBID, which should be retained if we carefully control the thermal budget during conversion to carbide.

Although morphological Ge quantum dots are not forming under optimized conditions for pattern retention, epitaxial electronic quantum dots may still form. Note that Volmer-Weber heteroepitaxial Ge quantum dots have been shown to grow on the basal plane of 4H-SiC[18] although we are not aware of similar studies on 3C-SiC. The lattice mismatch between Ge and SiC is -23%, which cannot be supported elastically even in a single monolayer. At such large mismatch, combined with the unfavorable Ge-C bond energy (no equilibrium GeC phase exists), it is reasonable to speculate that Ge is incommensurate with the underlying SiC, and largely relaxed to its bulk lattice parameter[19-21]. Ge grown on the Si in between the SiC sites, however, will be pseudomorphically strained to -4%. Therefore, a Si layer that is overgrown on the patterned regions will result in tensile strained Si over the Ge/SiC sites, but unstrained Si will grow over the areas in between, still providing a reduction in conduction band energy that will localize electrons above the template sites. Clearly, however, detailed microanalysis, e.g., by transmission electron microscopy, is needed to determine whether this scenario is valid. Preliminary I-V measurements of single and multi-dot structures on the nanodice pattern of Fig. 1, vs. regions with no QD, do yield clear differences in behavior[22].

## Conclusion:

In conclusion, we have shown that arbitrary placement control of SiC nanodots, with interdot spacing down to 22.5nm, a narrow size distribution, and sub-nm control over position, is achievable with EBID of ambient hydrocarbons. Careful control of cleaning and thermal budget is required to retain the narrow size distribution imposed by EBID. Excess carbon located on the Si surface between the patterned SiC sites has a strong retarding effect on Ge diffusion during MBE growth, which is critical to suppressing coarsening processes that destroy the narrow distribution. However, this also results in conformal growth of Ge across the patterned array, with no significant net accumulation at the template sites. Although no morphological Ge QDs are observed in this case, formation of electronic QDs could occur due to modulation of elastic strain going from growth on Si to growth on SiC. However, this remains to be definitively demonstrated through ongoing structural and transport investigations. At higher temperatures, Ge does begin to diffuse on the carbon-modified surface, resulting in formation of 3D Ge QDs localized at SiC sites, but coarsening also occurs. Analysis of the coarsened QD volumes suggests that a Ge wetting layer at least one monolayer thick is retained on the Si regions between the templated sites, presumably stabilized by surface energy minimization. Although the directed self-assembly mechanisms using this method are not as simple as previously envisioned, there is nonetheless strong potential for this straightforward approach to provide exchange-coupled QD arrays of arbitrary geometry, which would enable a number of novel spin-based interactions and operations to be investigated.

## Acknowledgments:

Support from the United States Department of Energy Office of Basic Energy Sciences is gratefully acknowledged under grant number: DE-FG02-07ER46421.